%% file: mxdag.tex
\PassOptionsToPackage{pdfpagelabels=false}{hyperref} 
\documentclass{hotnets17}

\usepackage{times}  
\usepackage{hyperref}
\usepackage[ruled,vlined]{algorithm2e}
\usepackage{multirow}

\hypersetup{pdfstartview=FitH, pdfpagelayout=SinglePage}

\usepackage{etoolbox}
\makeatletter
\patchcmd{\@maketitle}
  {\advance\dimen0 by -12.75pc\relax}
  {}
  {}
  {}

\newcommand{\mypara}[1]{\vspace{0.03in}\noindent\textbf{#1}\xspace}

\newcommand{\fref}[1]{\mbox{Fig.~\ref{#1}}}

\begin{document}



\title{MXDAG: A Hybrid Abstraction for Cluster Applications \vspace{-1em}}

\author{
\alignauthor Weitao Wang, Sushovan Das, Xinyu Crystal Wu, Zhuang Wang, Ang Chen, T. S. Eugene Ng\\
\affaddr Rice University}

\maketitle

\begin{abstract}
Distributed applications, such as database queries and distributed training, consist of both compute and network tasks.
DAG-based abstraction primarily targets compute tasks and has no explicit network-level scheduling.
In contrast, Coflow abstraction collectively schedules network flows among compute tasks but lacks the end-to-end view of the application DAG.
Because of the dependencies and interactions between these two types of tasks, it is sub-optimal to only consider one of them.
We argue that co-scheduling of both compute and network tasks can help applications towards the globally optimal end-to-end performance.
However, none of the existing abstractions can provide fine-grained information for co-scheduling.
We propose MXDAG, an abstraction to treat both compute and network tasks explicitly. 
It can capture the dependencies and interactions of both compute and network tasks leading to improved application performance.

\end{abstract}

\input{1-introduction}

\input{2-motivation}
\input{3-design}

\input{4-principle}

\input{5-usage}

\input{6-related}

\input{7-conclusion}

\section{Acknowledgement}

This research is sponsored by the NSF under CNS-1718980, CNS-1801884, and CNS-1815525.

\bibliographystyle{abbrv} 
\begin{small}
\bibliography{mxdag}
\end{small}

\end{document}

%% file: 1-introduction.tex
\section{Introduction} \label{sec:introduction}

Today's compute clusters host a variety of distributed applications ranging from map-reduce to distributed deep neural network training and database queries~\cite{vavilapalli2013apache, horovod, PSosdi2014, verma2015large}. 
These applications consist of several compute and communication stages. Due to the distributed nature of such applications running on emerging compute resources (CPU, GPU), the network can easily become the bottleneck~\cite{zhang2020network, bytescheduler, luo2018parameter}. Scheduling such applications directly impacts both the end-to-end application performance and cluster resource utilization. Directed Acyclic Graph (DAG) is the state-of-the-art abstraction for analyzing these cluster parallel jobs. Traditionally, the DAG-based scheduling systems, e.g., Spark \cite{zaharia2010spark}, Flink \cite{carbone2015apache}, Dryad \cite{isard2007dryad}, Tez \cite{saha2015apache} do not take the network resources into consideration, rather focus on the computational tasks and try to split those tasks onto different hosts. However, recent works on network-aware DAG scheduling \cite{grandl2016graphene,blagodurov2015multi,grandl2016altruistic} do take the network bandwidth resource into consideration and model the problem as resource packing. However, such strategy does not perform explicit network-level scheduling as the abstraction lacks fine-grained flow-level information. 

Another group of work focuses on explicit network scheduling and job placement where the primary objective is to localize most of the traffic flow between tasks and balance the network utilization across the cluster~\cite{dogar2014decentralized, corral,survive, sinbad, zimmer2016multi, tang2019spread, chowdhury2011managing}. Such frameworks have more fine-grained information about the network I/O, but lacks tight integration between network flows and application-level requirements. Coflow abstraction~\cite{chowdhury2012coflow} tries to bridge this gap by jointly considering collection of network flows among multiple compute stages, which enables application-aware network scheduling to some extent \cite{chowdhury2014efficient, chowdhury2015efficient}. However, coflow abstraction has several fundamental limitations that can lead to inefficient scheduling. Firstly, it does not have the global view of the application DAG. Secondly, defining a coflow inside an asymmetric DAG can be ambiguous. Finally, coflow treats the internal flows based on an all-or-nothing principle, which can obscure the critical path information and harm the application performance.

Therefore, co-scheduling of both compute and network tasks is necessary to improve the application performance and resource utilization. The fundamental benefit is that, co-scheduling would explicitly consider all kinds of dependencies from the application (compute-network, compute-compute, network-network) in a more fine-grained manner. Eventually, these relationships can be exploited to have better critical path analysis and achieve the optimal scheduling strategies e.g., overlapping data communication with computation, chunking up data flow to enable pipelining as necessary, deliberately preempting and ordering parallel traffic flows to accelerate downstream task execution etc.


 Hence, the key insight of co-scheduling is to treat both compute and network tasks explicitly. 
 Ideally, the co-scheduler can optimally schedule the application given the fine-grained information regarding both the compute and network tasks is available along with an end-to-end view.
 Unfortunately, none of the existing abstractions have the provision to encode such fine-grained information. This inherent gap between the ideal co-scheduling requirements and existing abstractions motivates us to propose a more general and fundamentally different abstraction, called MXDAG. It abstracts both the compute and network tasks in a DAG as explicit nodes annotated with fine-grained information. The arrows capture all kinds of dependencies (compute-network, compute-compute, network-network). MXDAG can potentially address several challenges related to co-scheduling system design.  
Firstly, by decoupling compute and network tasks, MXDAG enables the co-scheduler to treat them uniquely as they have fundamentally different behavior. On one hand, compute tasks can be easily isolated among CPU/GPU cores but the performance is less predictable. On the other hand, network tasks are more predictable given the data size and network bandwidth are known, but they cannot be isolated so easily.
Secondly, equipped with more fine-grained information, MXDAG enables the co-scheduler to consider the heterogeneity in both compute (CPU, GPU) and network resources arising from technology advancements \cite{narayanan2020heterogeneity, jiang2020unified, park2020hetpipe}.  Finally, with global view of the application, MXDAG enables the co-scheduler to carefully analyze the impact of pipelining between all kinds of tasks and make smart decisions.  
 
 
The paper is organized as follows. In Section \ref{sec:motivation}, we discuss the limitation of existing abstractions (DAG and Coflow) and provide complete analysis of pipelineability. In Section \ref{sec:definition}, we formally define our abstraction MXDAG in detail. Section \ref{sec:schedule} describes the key principles regarding MXDAG-based scheduling with several use cases. We discuss related work in Section~\ref{sec:related} and conclude in Section~\ref{sec:conclusion}.

%% file: 2-motivation.tex
\section{Motivation} \label{sec:motivation}

\begin{figure}[t!]
    \centering
      \includegraphics[width=0.45\textwidth]{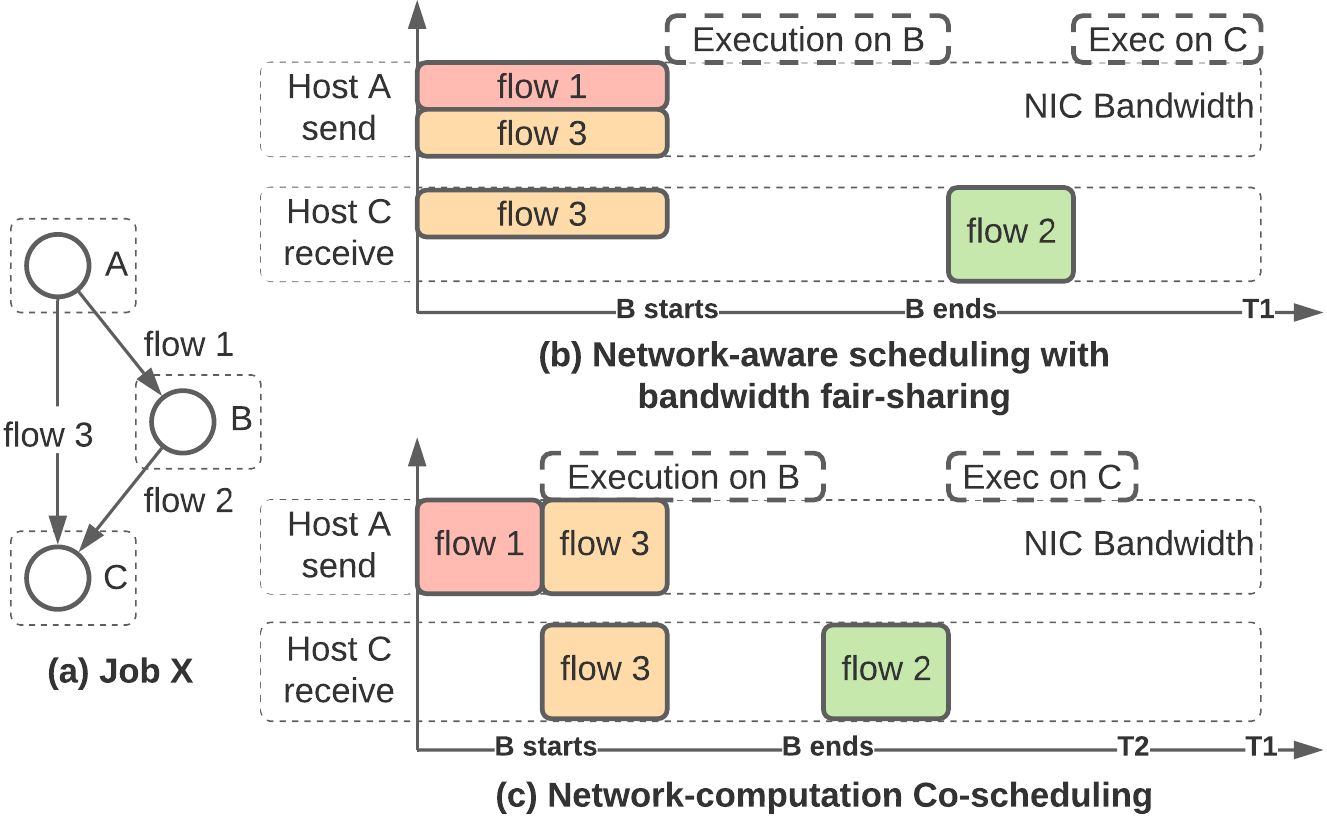}
      \vspace{-3mm}
      \caption{Comparison between network-aware scheduling and network-compute co-scheduling.}
      \vspace{-3mm}
      \label{fig:dag_moti}
\end{figure}


\subsection{Previous DAGs Lack Explicit Network Flow Scheduling}
DAG-based abstraction is widely used to analyze and optimize parallel jobs in computing clusters. It could capture data flows between computational tasks and indicate corresponding dependencies, which plays a significant role in resource sharing and task scheduling. However, most of the existing DAG-based scheduling frameworks, including   
Spark~\cite{zaharia2010spark}, Flink~\cite{carbone2015apache}, Dryad~\cite{isard2007dryad}, Tez~\cite{saha2015apache}, mainly focus on the host-level computational tasks and implicitly treat the network requirements as parts of the computational tasks. Several recent network-aware DAG-based schedulers \cite{grandl2016graphene,blagodurov2015multi,grandl2016altruistic} start to take the bandwidth resource into consideration, 
Nevertheless, they only consider the bandwidth when packing different resources while no explicit flow-level resource scheduling information is included. Therefore, these DAG abstractions usually use the same type of edges, without thoroughly distinguishing between logical dependencies and real data transmissions, which could lead to inefficient scheduling results.


For example in \fref{fig:dag_moti}(a), host A needs to send two flows to host B and C. A network-aware scheduler would fairly share the bandwidth resources and schedule the tasks as in \fref{fig:dag_moti}(b), flow 1 and flow 3 will share the NIC bandwidth and thus extend the completion time. As a result, job X can only complete at time T1. Instead, a network-computation co-scheduler would schedule in a globally optimal way by prioritizing the flow 1 over flow 3 as \fref{fig:dag_moti}(c), so that both flow 1 and 3 can enjoy the full bandwidth. Therefore, the task on C could start at time T2, much earlier than the previous case. 


\begin{figure*}[t!]
    \centering
      \includegraphics[width=0.95\textwidth]{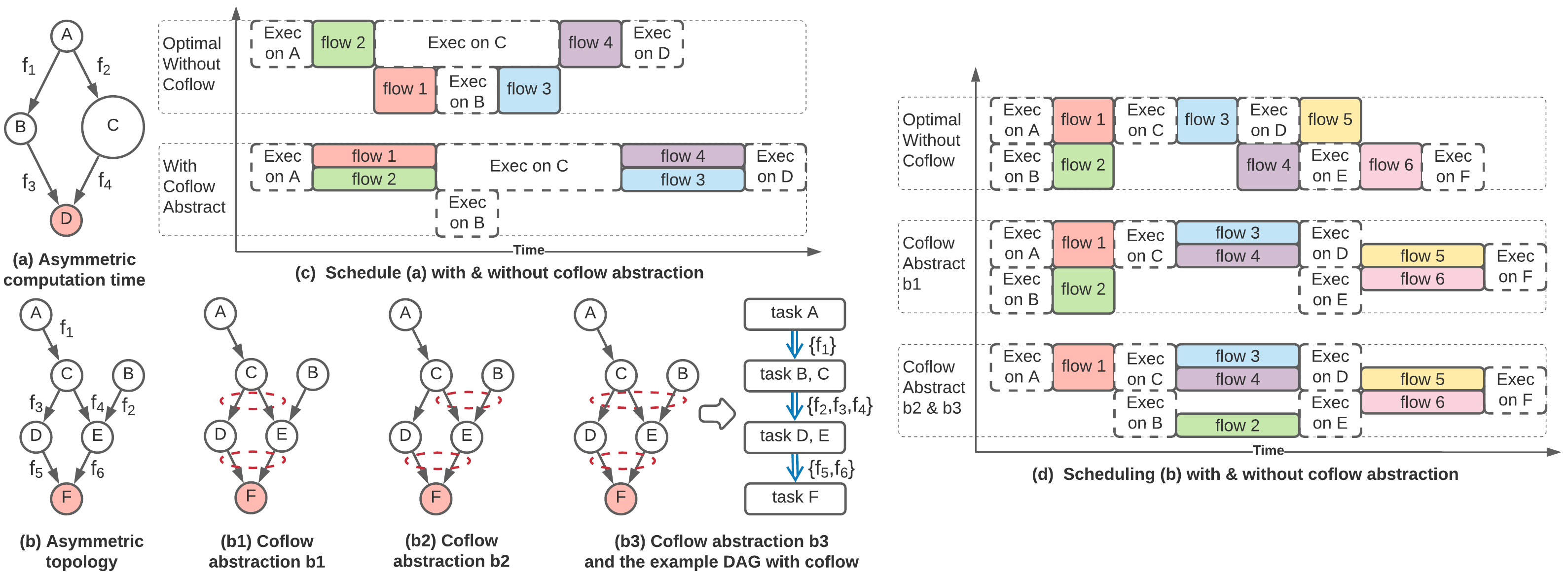}
      \vspace{-3mm}
      \caption{(a) DAG with asymmetric computation times; (b) DAG with asymmetric topology from Wukong~\protect\cite{carver2020wukong}; (b1,b2,b3) Three potential coflow abstractions of DAG in (b), and flows grouped with dashed ellipse are considered as a coflow; (c) An optimal schedule without coflow for DAG in (a) and the schedule with $\{f_1, f_2\}$ and $\{f_3, f_4\}$ as coflows; (d) An optimal schedule without coflow abstraction for DAG in (b) and schedules with different coflow abstraction.}
    \label{fig:coflow_compare}
\end{figure*}

\begin{figure*}[t!]
    \centering
      \includegraphics[width=\textwidth]{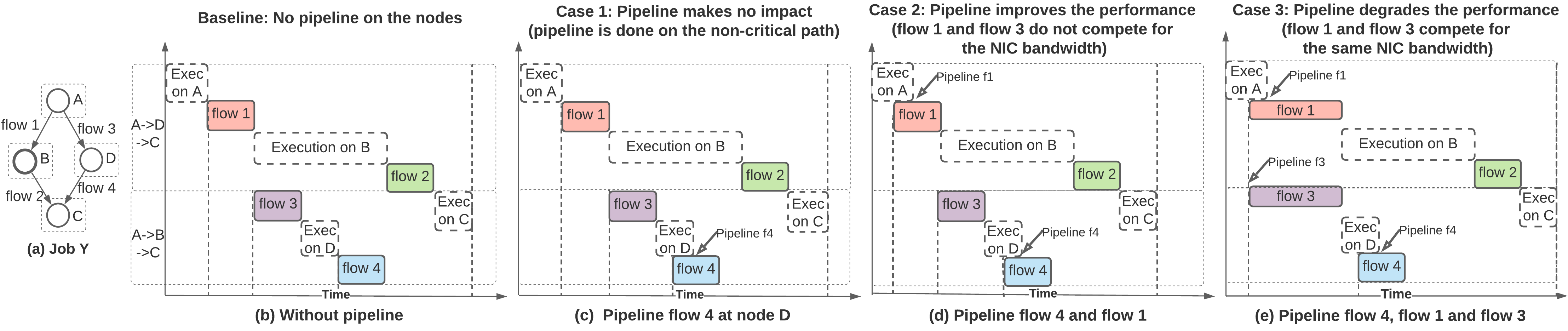}
      \vspace{-7mm}
      \caption{Different pipelineablity choices make different impacts on application performance}
      \vspace{-3mm}
      \label{fig:pipelineability} 
\end{figure*}

\subsection{Coflow Abstraction Lacks Global View}

The Coflow abstraction~\cite{chowdhury2012coflow} is proposed a decade ago and is widely used by many network schedulers to optimize resource sharing~\cite{chowdhury2014efficient, chowdhury2015efficient, li2016efficient}. Such abstraction jointly considers parallel flows between two subsets of hosts having a common objective \cite{chowdhury2012coflow} and also contains the information about the communication pattern e.g., broadcast (one-to-many), aggregation (many-to-one), shuffle (many-to-many), etc. However, coflow abstraction has two fundamental limitations:

\textbf{Coflow abstraction implicitly assumes symmetry in the DAG, which leads to definitional ambiguity when abstracting asymmetric DAG.} Asymmetric DAG is common for several cluster applications~\cite{carver2020wukong, lopez2020triggerflow, wu2015hierarchical, shankar2018numpywren}. From those, we primarily observe two sources of asymmetry:
    1) For a DAG with a symmetric structure, the asymmetry can arise from the heterogeneity in computation time across the nodes, as shown in ~\fref{fig:coflow_compare}(a). The computation times for tasks on host B and C can be unequal ($t_1$ and $t_2$ respectively) due to the heterogeneity in the underlining hardware (GPU, CPU, etc.) or the different task sizes; 
    2) The DAG can also have an asymmetric topology as shown in ~\fref{fig:coflow_compare}(b) (adopted from \cite{carver2020wukong}).
        From which, three different coflow abstractions might be derived as ~\fref{fig:coflow_compare}(b1,b2,b3). 
            In~\fref{fig:coflow_compare}(b1), we consider two coflows i.e., broadcast from node C ($f_3$ and $f_4$) and aggregation at node F ($f_5$ and $f_6$).
            In~\fref{fig:coflow_compare}(b2), aggregation at node E ($f_2$ and $f_4$) is considered to be an alternative coflow.
            In~\fref{fig:coflow_compare}(b3), all flows between nodes \{B,C\} and \{D,E\} are considered to be one coflow ($f_2$, $f_3$ and $f_4$). 
        Despite having several options for defining a coflow, the application programmer must commit to a specific definition while writing the application. It cannot be modified during runtime. Most importantly it is difficult to predict, how a specific coflow definition would impact the application performance. 

\textbf{Without a global view of the DAG, the coflow abstraction could lead to inefficient scheduling.} 
By enforcing all the flows in a coflow to end at the same time, coflow can possibly obscure the critical path information of the DAG, which may lead to sub-optimal performance during scheduling, as shown in ~\fref{fig:coflow_compare}(c) and (d). 
    For the DAG with asymmetric computation time, 
        the optimal scheduling without coflow in ~\fref{fig:coflow_compare}(c) treats each flow individually, and allows each flow to avoid sharing the NIC bandwidth resources smartly. 
        While with the coflow abstraction, coflow $\{f_1,f_2\}$ and  $\{f_3,f_4\}$ have to share the NIC bandwidth at the same time and enlarge the end-to-end completion time.
    For the DAG with asymmetric topology,
        The optimal solution delays the start time for $f_4$ and avoid resource sharing on the source NIC on the host C. And as a cascading effect, the $f_5$ and $f_6$ also do not share the bandwidth on the destination NIC on the host F.
        Whereas, the three different solutions with coflow abstraction all have sub-optimal scheduling. 
            In ~\fref{fig:coflow_compare}(d), the coflow abstraction b1 force the coflows \{$f_3$, $f_4$\} and  \{$f_5$, $f_6$\} to share the NIC bandwidth on hosts C and F respectively, so that the execution on D will be postponed. 
            Meanwhile, the coflow abstractions b2 and b3 also forces the scheduler to schedule $f_2$ and $f_4$ together as one coflow, but lead to the NIC bandwidth competition of host E.

\subsection{Both DAG and Coflow Abstractions Lack Pipelineability Analysis}

Pipelining is a promising strategy to improve the performance of distributed applications. By chunking up the data flows, not only the storage usage on the host can be reduced, but also the overlap between communication with computation can be maximized. There are two common kinds of applications that could be optimally scheduled with efficient pipelining. On one hand, map-reduce jobs could significantly reduce the job completion time by pipelining the execution of the map and reduce tasks~\cite{condie2010mapreduce}. On the other hand, in distributed deep learning, especially the gradient aggregation part can benefit a lot from pipelining the push and pull operations, thereby significantly reducing the communication time and accelerating the overall training~\cite{peng2019generic}.

However, none of the existing DAG-based and coflow-based abstractions fully consider the pipelineability in their scheduling strategies ~\cite{grandl2016graphene, chowdhury2012coflow}. Caerus~\cite{zhang2021caerus} does consider pipelineability and provides a step dependency model to capture the pipeline information. Nevertheless, it only profiles the pipelineability on the computational tasks, without any network-level pipelining analysis. Therefore, such network-oblivious pipelining could lead to sub-optimal scheduling decisions.  

We analyze several situations where pipelineablity has different impacts, using a four-node DAG with A->B->C as the critical path, as shown in \fref{fig:pipelineability}(a). \fref{fig:pipelineability}(b) displays the execution timeline of the baseline situation where pipelines are not allowed anywhere. We then illustrate three different scenarios with different pipelineablity choices as follows to provide the insights of pipelining impacts. With these insights, we could observe that a better scheduler should allow network operators to choose whether to use pipeline or not and which tasks need to be pipelined at runtime.

\mypara{Case 1: Pipelining on the non-critical path makes no impact on the application performance.} As shown in the \fref{fig:pipelineability}(c), pipelining flow 4 on node D will not affect the length of the
critical path since node D does not belong to the critical path. Therefore, the execution on C will be the same as the baseline case and there are no changes to the overall application performance. 

\mypara{Case 2: Pipelining on the critical path can improve the application performance.} As shown in \fref{fig:pipelineability}(d), besides pipelining flow 4 as the previous case, flow 1 on node A is also chosen to be pipelined. Since flow 3 still starts after flow 1 is completed, these two flows will not overlap and 
enjoy the full NIC bandwidth of node A.
We observe that such pipeline strategy will shorten the critical path length (A->B->C), causing the execution on node C to start earlier than in the baseline. 

\mypara{Case 3: Pipelining on the critical path can degrade the application performance.} As shown in \fref{fig:pipelineability}(e), 
allowing to pipeline flow 3 along with flow 1 and flow 4, will increase the critical path length.
In this case, flow 1 and flow 3 start at the same time and take twice the time to finish as they share the same NIC bandwidth of node A. Therefore, the length of the critical path (A->B->C) becomes longer that causes the execution on node C to start later than in the baseline. 

%% file: 3-design.tex
\section{MXDAG} \label{sec:definition}

To address the above drawbacks of the existing solutions, we introduce the MXDAG abstraction.
The construction of MXDAG can still rely on existing solutions to get the necessary information for different kinds of applications. On one hand, there are bare-metal applications like distributed deep learning and distributed matrix computation, where all the necessary information (e.g., CPU/GPU cores, data size, NIC bandwidth) can be provided explicitly before execution. On the other hand, for the applications running on Spark, Hadoop, Dryad etc., such information are not known {\em a priori}, because the physical placement is decided by system-schedulers (e.g., YARN) at runtime. For these cases, flow-level details can be estimated from historical placement information~\cite{zhang2021caerus, peng2018optimus, gu2019tiresias} and execution time of a compute task on specific hardware can be estimated by measurement-based job profiling \cite{wyatt2018prionn,sonmez2009trace,zhang2015prism,wasi2016comprehensive}. 

\begin{figure}[t!]
    \centering
      \includegraphics[width=0.47\textwidth]{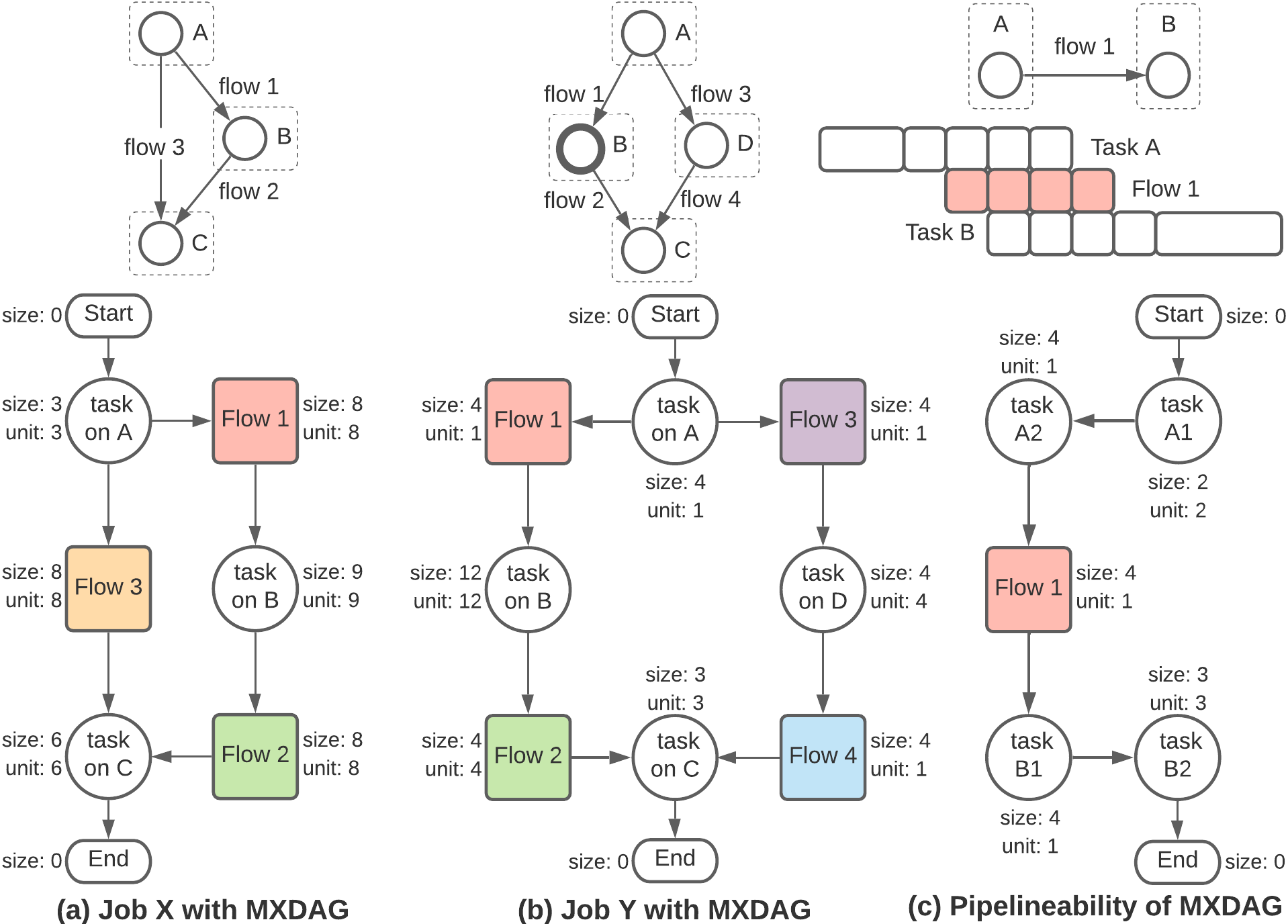}
      \vspace{-1mm}
      \caption{(a, b) MXDAGs for Job X and Y; (c) MXDAG with only partial parts of Task A and Task B being pipelined.}
      \vspace{-4mm}
      \label{fig:mxdag}
\end{figure}

\subsection{Definition}
\mypara{MXTasks} serve as the nodes $\{v_1, v_2, ..., v_k\}$ in the MXDAG $G$, and each MXTask represents either a task running on a host using CPU/GPU/Accelerator or a flow in the network with a single sender and receiver. Note that all the MXTasks are physical processes or flows, rather than logical tasks which usually contain multiple physical tasks on multiple machines. 
    Each MXTask is a procedure which receives an input and gives an output after being processed by a certain amount of resources. 
    To include more quantitative information, each MXTask has two additional fields:
    1) MXTask size $Size(v_i)$ represents the completion time of a MXTask with the maximum resource assigned (computation or bandwidth), which has similar meaning with the concept of task durations in Decima~\cite{mao2019learning} and Graphene~\cite{grandl2016graphene}. And the size can be used to estimate the completion time when only partial resources are assigned to the task; 
    2) MXTask unit $Unit(v_i)$ represents the size of the smallest unit when being executed under the pipeline. 
Note that for MXTasks that cannot be executed in a pipeline, its unit size is equal to its task size. 

\mypara{MXDAG} is a directed graph $G=(V,E)$ composed of MXTasks $V=\{v_S, v_1, v_2, ..., v_k, v_E\}$ and dependencies represented as $E=\{e_1, e_2, ..., e_i\}$. 
$v_S$ and $v_E$ are the dummy start and end MXTasks in an MXDAG.
An edge from $v_i$ to $v_j$ indicates that task $v_j$ cannot start before $v_i$ ends. MXDAG serves as an abstraction for a cluster application or an individual function within an application, the latter being very common in serverless environments. For instance, the MXDAGs for job X and Y are shown in \fref{fig:mxdag}(a) and (b).
Different from existing DAG-based systems, MXDAG elevates the network flows to the same level as the computational tasks on the hosts. Therefore, MXDAG can provide detailed information as well as the importance of each network flow, figuring out the relative priorities and achieving better scheduling strategies. 

\begin{figure}[t!]
    \centering
      \includegraphics[width=0.34\textwidth]{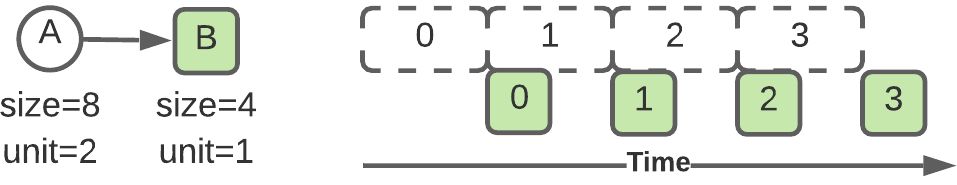}
      \vspace{-3mm}
      \caption{Example Pipeline for two pipelineable MXTasks with different task sizes and task unit sizes.}
      \vspace{-10mm}
      \label{fig:pipeline}
\end{figure}

\mypara{Pipelineability.}
To include the pipelineability of MXTasks, pipelineable MXTask divides its input and output into minimum units. Namely, once an input unit is received, that MXTask can start processing and immediately give an output unit as shown in \fref{fig:pipeline}.
    For the network MXTasks, as long as the output of the predecessor CPU task can be given in units, the pipelineability can be enabled instead of waiting until all outputs are ready (E.g., sending serialized objects like hash tables). 
    While for the computational tasks, we rely on the existing pipeline analysis works, like the per-step dependency model in Caerus~\cite{zhang2021caerus}, to profile the pipelineability in the computational tasks. For the computational tasks with both pipelineable part and sequential-execution-only part, two MXTasks will be derived as the task A and B in \fref{fig:mxdag}(c). 


\subsection{Properties of MXDAG}

Firstly, we will introduce several notations and properties of MXDAG that are useful for the following discussions. 
    \textit{Path} in the MXDAG denotes a finite sequence of edges which join a sequence of MXTasks with a head task (node) and a tail task. 
    \textit{Copath} denotes a group of paths with the same head node and tail node, as the path $A$->$f_1$->$B$->$f_2$->$C$ and $A$->$f_3$->$C$ in job X of \fref{fig:mxdag}(a). 
    The \textit{Path Length}, representing the end-to-end computation time for a path, is calculated recursively in an MXDAG: 1) divide a path into Copaths and normal paths without Copath, and its length is treated as the sum of normal path lengths and Copath lengths; 2) For a Copath, its length is equal to the length of its longest path; 3) For a normal path, its length can be calculated as the sum of the pipelineable-only paths and sequential-only paths. The length of a sequential-only path $P_{seq}=\{v_0, v_1, ..., v_m\}$ and a pipelineable-only path $P_{pipe}=\{v_0, v_1, ..., v_n\}$ can be calculated as below (Given the resource assigned to each MXTask $v_i$ as $Rsrc(v_i)$):

\vspace{-3mm}

\begin{equation}
\begin{aligned}
    \small
    Len(P_{seq}) &= \sum_{i=0}^m{\frac{Size(v_i)}{Rsrc(v_i)}}
\end{aligned}
\end{equation}

\vspace{-3mm}

\begin{equation}
\begin{aligned}
    \small
    Len(P_{pipe}) 
    =& \sum_{i=0}^n{\frac{Unit(v_i)}{Rsrc(v_i)}} + \max_i\{\frac{Size(v_i)}{Rsrc(v_i)}\} \\
    & - \max_i\{\frac{Unit(v_i)}{Rsrc(v_i)}\}
\end{aligned}
\end{equation}

The equation (2) implies that the length of a pipelineable-only path is dominated by the pipelineable task with the longest execution time as shown in \fref{fig:pipeline}. Moreover, we can observe that the maximum throughput of the flow can also be restricted by the CPU processing speed when pipeline is used. 


Another important property is that the paths within any Copath have the same barriers, so that every Copath has a critical path. With these barriers, only all of the paths within that Copath have finished execution, or given the first unit of result in a pipeline, that tail node can start the execution. We define the path with the longest length in a Copath as its critical path, then the length of the critical path determined the overall execution time of a Copath.

%% file: 4-principle.tex
\section{Using MXDAG} \label{sec:schedule}




\subsection{Schedule A Single MXDAG}

\mypara{Objective: } 
Single MXDAG scheduling aims to minimize the job completion time (JCT) considering both computational and network tasks. The collective objective of all the paths in the MXDAG can be expressed as

\vspace{-4mm}

\begin{equation}
\begin{aligned}
    &\min \max_{P \in \mathcal{P}}\{Len(P)\} \\
    \text{where } \mathcal{P}=\{P&|Head(P)=v_S, Tail(P)=v_E\}
\end{aligned}
\end{equation}

To achieve the above objective, we will use MXDAG to analyse the inherent dependencies and resource sharing between the MXTasks. Since the optimal scheduling for MXDAG is an NP-hard problem~\cite{grandl2016graphene, feige2002improved, mastrolilli2008acyclic}, we will give several principles to guide the scheduling and inspire new heuristics, leaving the detailed algorithm as future works. 


\mypara{Principle 1: Prioritize the critical path over non-critical paths within any Copath, without letting the non-critical paths have longer completion time than the critical path.}

If the different paths within a Copath share some resources, like the NIC bandwidth or the CPU cores, delaying the resource allocation for the non-critical paths or allocating fewer resources to the non-critical paths could help shrink the critical path completion time. By ensuring the non-critical paths has shorter or equal completion time with the critical path, the over-optimization can be avoided. 
    Notably, though the pipeline can be used to shrink the delay between two tasks, it also enforces the resources to be occupied right after the precedent task begin processing, which may contend with the tasks on the critical path. Thus, even for pipelineable MXTasks, the pipelines will only be applied when they can shrink the overall execution time.
    
\subsubsection{Example Case: Distributed Deep Learning}

\begin{figure}[t!]
    \centering
      \includegraphics[width=0.47\textwidth]{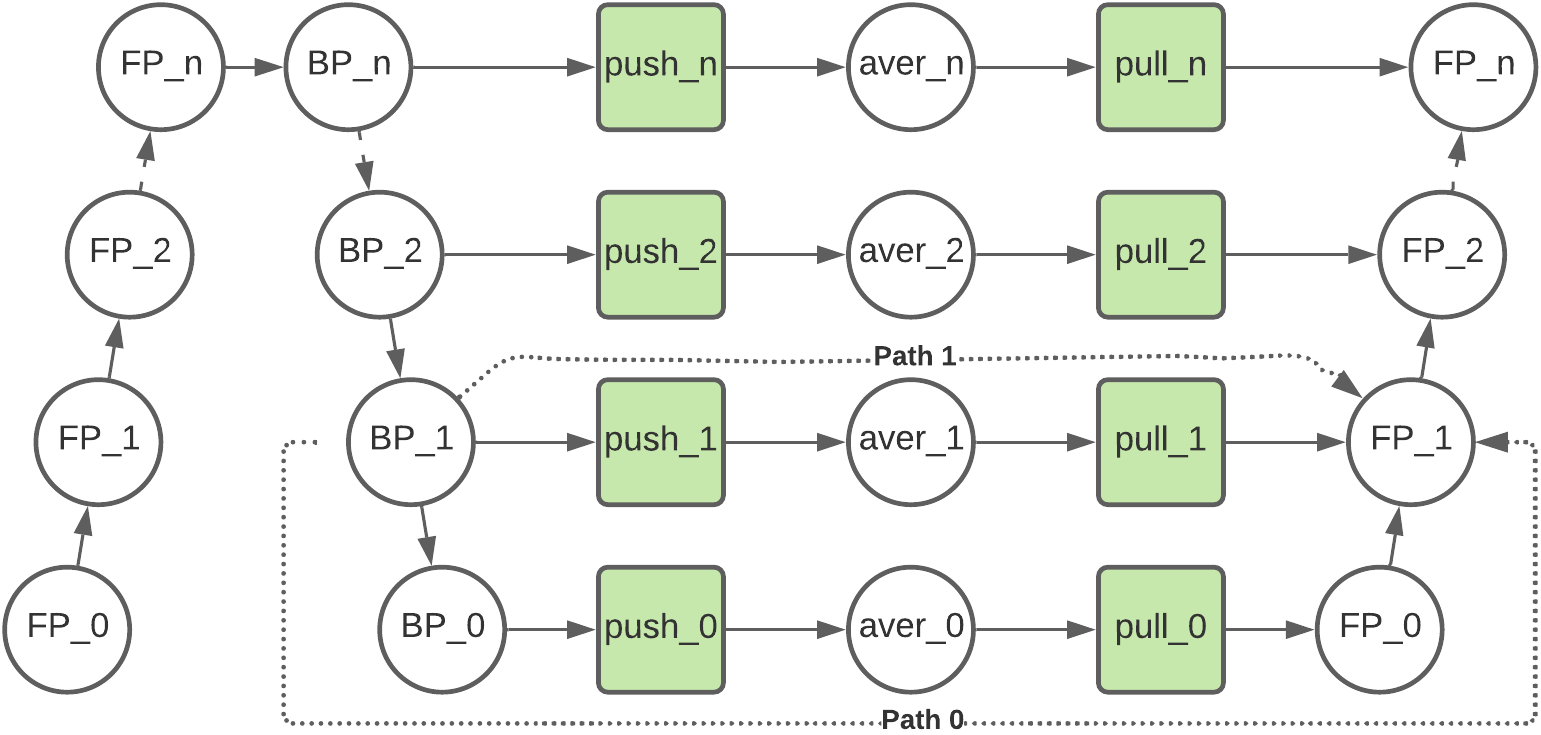}
      \vspace{-2mm}
      \caption{MXDAG for Distributed Machine Learning}
      \vspace{-4mm}
      \label{fig:bytescheduler}
\end{figure}

We take a widely-used and increasing-important application for example --- Data-parallel Distributed Deep Learning.
    The communication overhead of synchronizing the parameters on different machines is significant to the data-parallel distributed learning workloads, but transmit the parameter in layers can help shrink the overall completion time. To explain the idea of layer-wise parameter synchronization, \fref{fig:bytescheduler} shows the MXDAG for that process: the parameters of each layer are generated after the back-propagation (BP) process on the GPU, and the synchronized parameters will be used by the forward-propagation (FP) process in the next iteration. For neural networks with multiple layers, the FP processes are executed from lower layer to higher layer, and the BP processes are executed in a reverse manner.  

Take the path 0 and path 1 for example, the MXTasks $push_i$ and $pull_i (i\in \{0,1\})$ share the same bandwidth resource over the network. Consider the Copath between $BP_1$ and $FP_1$, if path 0 is the critical path, then all the bandwidth resource should be allocated to the path 0 to achieve the shortest completion time. Whereas, if the path 1 is the critical path, as long as the $FP_0$ finishes earlier than the $pull_1$, there is no strict ordering for the resource allocation between path 0 and path 1. (Note that strictly prioritizing the path 0 is an optimal scheduling within the above solution space.) ByteScheduler~\cite{peng2019generic} rearranged the tensor transmission order for each layer to accelerate the training process. Their solution decide to strictly prioritize the parameters in the lower layers over the parameters in the upper layers, while our analysis over MXDAG echos their solution. 

\subsection{Schedule Multiple MXDAGs}

Besides minimizing the JCT, scheduling multiple MXDAGs has more objectives, like meeting the deadline of each job and ensuring fairness among all the MXDAGs.
Since the key of multiple MXDAG scheduling is also resource sharing, here we give our second principle to guide the resource allocation over multiple MXDAGs. 

\mypara{Principle 2: Let each MXDAG to be altruistic by delaying its non-critical path resource allocation to benefit other MXDAG's critical path, without increasing its own end-to-end completion time.} 

Once the resource allocation for the critical path has been determined insides the MXDAG, the overall end-to-end job completion time is certainly bounded with the execution time of the critical path. Thus, as long as other MXTasks on the non-critical path finishes earlier than the critical path, the shortest JCT is preserved. With this idea, we allow the scheduler to delay the resource allocation for those non-critical MXTasks, since the resources saved during those waiting time, can be allocated to other application's critical MXTasks for a shorter JCT.

\subsubsection{Example Case: Map-reduce Applications}

\begin{figure}[t!]
    \centering
      \includegraphics[width=0.47\textwidth]{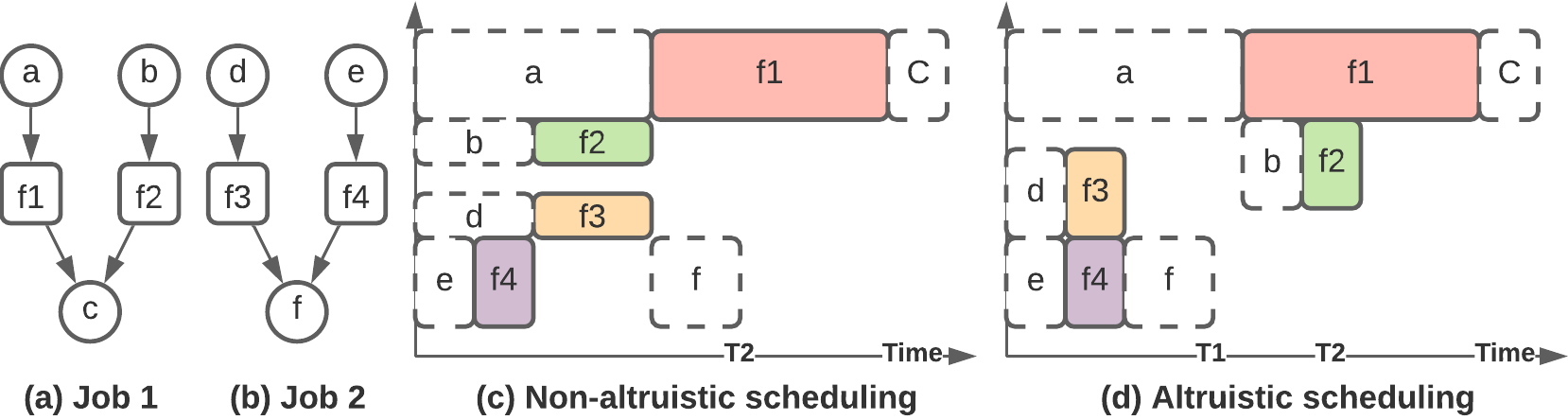}
      \vspace{-3mm}
      \caption{Schedule multiple Map-Reduce Jobs. MXTask b and d share the same computational resource and MXTask f2 and f3 share the same NIC bandwidth resource.}
      \vspace{-3mm}
      \label{fig:mapreduce}
\end{figure}

For the example map-reduce jobs in \fref{fig:mapreduce}, MXTask $a$ and $f_1$ have longer computation time than MXTask $b$ and $f_2$. While the MXTask $d$ and $f_3$ from job 2 share the same computational / bandwidth resource with $b$ and $f_2$ from job 1. 

Without altruistic scheduling, task $d$ and $b$, $f_2$ and $f_3$ will share the resource lead to a longer completion time for job 2 in \fref{fig:mapreduce}(c). However in \fref{fig:mapreduce}(d), with principle 2, though job 1 cannot benefit itself from delaying the resource allocation for $b$ and $f_2$, its altruistic behavior helps the job 2 to finish earlier from T2 to T1 by shrinking the ciritical path in the job 2 MXDAG. This scheduling plan is also compatible with another job-scheduling work's solution --- CARBYNE~\cite{grandl2016altruistic}.

%% file: 5-usage.tex
\subsection{Other Usages} \label{sec:usage}


\mypara{What-if analysis on cluster applications.}
MXDAG can be used to conduct a what-if analysis on the cluster applications, including whether to pipeline compute and network tasks, whether to re-partition work among compute and network tasks, which are not possible with traditional DAG. For instance, a cluster application developer can use the MXDAG of an application to determine whether a revised application design that enables pipelining between previously non-pipelined MXTasks can help shrink the end-to-end completion time.

\mypara{Monitoring and debugging cluster applications}
The estimated task execution time may be different from the actual execution time due to inaccurate data or unexpected events during runtime. By monitoring the progress of each path and the barriers in MXDAG, we can efficiently and accurately identify the unexpected events and the corresponding host straggler or network straggler, while traditional DAG cannot distinguish those two kinds of stragglers. Moreover, operators could leverage the current progress and determine the new critical paths to optimize the scheduling plan at runtime. 





%% file: 6-related.tex
\vspace{-3mm}
\section{Related work}
\label{sec:related}

\textbf{Network-aware DAG.} Network-aware DAG schedulers e.g., Graphene \cite{grandl2016graphene}, Tetris \cite{blagodurov2015multi}, Network machine \cite{giroire2019network} modify the DAG abstraction to treat network bandwidth as a dividable resource and provide greedy heuristics to efficiently pack the tasks. There is no explicit scheduling of network flows due to the lack of flow-level information.

\textbf{Coflow.} Explicit network schedulers e.g., Baraat \cite{dogar2014decentralized}, Varys \cite{chowdhury2014efficient}, Aalo \cite{chowdhury2015efficient} fundamentally consider Coflow abstraction and perform application-aware network scheduling. Although such network schedulers explicitly schedule the network flows, they lack the global view of the application DAG which makes the critical path information elusive.

\textbf{DAG + Coflow.} Recent work like Branch-scheduling \cite{hu2019branch} extends the traditional DAG abstraction and glue that with Coflow. Although such extended abstraction provides a slightly better way to capture the compute-network dependencies, the fundamental limitations of both DAG and Coflow abstractions still remain. On one hand, it does not decouple compute and network tasks explicitly. On the other hand, there is no provision of prioritizing the flows inside a coflow which could potentially benefit the application. 

In contrast, MXDAG decouples the compute and network tasks, captures dependencies in a more fine-grained manner with an end-to-end application view and enables explicit co-scheduling. As our abstraction can better characterize applications, it can potentially benefit more recently proposed deep neural network scheduler such as Decima~\cite{mao2019learning}.

%% file: 7-conclusion.tex
\vspace{-3mm}
\section{Conclusion}
\label{sec:conclusion}
We studied the limitations of both DAG-based and coflow-based abstraction for resource scheduling in distributed applications.
We proposed an abstraction called MXDAG to express the dependencies and interactions between compute and network tasks and provide fine-grained information for the co-scheduling of both compute and network resources.